\documentclass[12pt]{article}

\usepackage{xcolor}             
\usepackage{amsmath,amssymb} 
\usepackage{graphicx} 
\usepackage{authblk}
\usepackage{caption}
\usepackage{titlesec}
\usepackage{float}
\usepackage{cite}
\usepackage{hyperref}
\usepackage{parskip}
\hypersetup{
    hidelinks
}

\usepackage{longtable}
\usepackage{booktabs}
\usepackage{array}

\usepackage{lmodern}
\usepackage[T1]{fontenc}
\usepackage[utf8]{inputenc}
\usepackage{parskip}

\title{\LARGE \textbf{DrugGen: Advancing Drug Discovery with Large Language Models and Reinforcement Learning Feedback}}
\author[1,2]{\large Mahsa Sheikholeslami}
\author[1]{\large Navid Mazrouei}
\author[1,3]{\large Yousof Gheisari}
\author[2]{\large Afshin Fasihi}
\author[1,4*]{\large Matin Irajpour}
\author[1,5*]{\large Ali Motahharynia}

\affil[1]{\small Regenerative Medicine Research Center, Isfahan University of Medical Sciences, Isfahan, Iran}
\affil[2]{\small Department of Medicinal Chemistry, School of Pharmacy, Isfahan University of Medical Sciences, Isfahan, Iran}
\affil[3]{\small Department of Genetics and Molecular Biology, Isfahan University of Medical Sciences, Isfahan, Iran}
\affil[4]{\small Isfahan Cardiovascular Research Center, Cardiovascular Research Institute, Isfahan University of Medical Sciences, Isfahan, Iran}
\affil[5]{\vspace{1em}\small Isfahan Neuroscience Research Center, Isfahan University of Medical Sciences, Isfahan, Iran}
\affil[*]{\small \textbf{Corresponding authors:} \

Matin Irajpour (\href{mailto:2012irajpour@gmail.com}{2012irajpour@gmail.com}; ORCID: 0000-0001-8504-9652)\newline
Ali Motahharynia (\href{mailto:alimotahharynia@gmail.com}{alimotahharynia@gmail.com}; ORCID: 0000-0002-1140-3257; Tel: +98 313 668 7087)}

\usepackage[margin=1in, top=1in, bottom=1in]{geometry}
\titleformat{\section}
  {\normalfont\Large\bfseries}
  {\thesection.}
  {0.5em}
  {}
\titleformat{\subsection}
  {\normalfont\large\bfseries}
  {\thesubsection.}
  {0.5em}
  {}
\titleformat{\subsubsection}
  {\normalfont\large\bfseries}
  {\thesubsubsection.}
  {0.5em}
  {}
\newcommand{\subsubsubsection}[1]{
  \paragraph{#1}
  \mbox{}
}
  
\begin{document}

\maketitle

% Abstract
\begin{abstract}
\noindent Traditional drug design faces significant challenges due to inherent chemical and biological complexities, often resulting in high failure rates in clinical trials. Deep learning advancements, particularly generative models, offer potential solutions to these challenges. One promising algorithm is DrugGPT, a transformer-based model, that generates small molecules for input protein sequences. Although promising, it generates both chemically valid and invalid structures and does not incorporate the features of approved drugs, resulting in time-consuming and inefficient drug discovery. To address these issues, we introduce DrugGen, an enhanced model based on the DrugGPT structure. DrugGen is fine-tuned on approved drug-target interactions and optimized with proximal policy optimization. By giving reward feedback from protein-ligand binding affinity prediction using pre-trained transformers (PLAPT) and a customized invalid structure assessor, DrugGen significantly improves performance. Evaluation across multiple targets demonstrated that DrugGen achieves 100\% valid structure generation compared to 95.5\% with DrugGPT and produced molecules with higher predicted binding affinities (7.22 {[}6.30-8.07{]}) compared to DrugGPT (5.81 {[}4.97-6.63{]}) while maintaining diversity and novelty. Docking simulations further validate its ability to generate molecules targeting binding sites effectively. For example, in the case of fatty acid-binding protein 5 (FABP5), DrugGen generated molecules with superior docking scores (FABP5/11, -9.537 and FABP5/5, -8.399) compared to the reference molecule (Palmitic acid, -6.177). Beyond lead compound generation, DrugGen also shows potential for drug repositioning and creating novel pharmacophores for existing targets. By producing high-quality small molecules, DrugGen provides a high-performance medium for advancing pharmaceutical research and drug discovery.
\end{abstract}

% Keywords Section
\textbf{Keywords:} Drug design; Drug repurposing; Large language model; Reinforcement learning; Molecular docking

% Introduction Section
\section{Introduction}\label{introduction}
Traditional drug design often falls short in handling the vast chemical
and biological space features involved in ligand-receptor interactions \cite{Bai2024, Coley2020}. Usually, a major proportion of suggested drug candidates fail in clinical trials \cite{Sun2022}, making drug discovery a time-consuming
and costly process. Recent advances in deep learning (DL), particularly
in generative models, offer promising solutions for these obstacles \cite{Tong2021, Zeng2022}. Deep learning models have been extensively used in molecular design \cite{meyers2021, Mendez_Lucio2020}, pharmacokinetics \cite{Janssen2024, Ota2022, Horne2024, Ghayoor2024}, pharmacodynamics predictions \cite{Menke2021}, and toxicity assessments \cite{Horne2024}. These models
improve the efficiency and accuracy of various tasks in drug
development, contributing to different stages of drug discovery and
optimization projects \cite{Qureshi2023, Zhuang2021}. However, due to the insufficiency of
available datasets, complexity of drug-target interactions, and
complication of manipulating complex chemical structures, generative DL
models also seem to be insufficient in proposing optimal answers to drug
design problems \cite{Gangwal2024}. Nevertheless, with the advancement of
transformer-based architecture in large language models (LLMs), new
horizons have opened up in various biological contexts. ProGen, a model
developed to design new proteins with desired functionality and
protein-ligand binding affinity prediction using pre-trained
transformers (PLAPT), a model for protein-ligand binding affinity
prediction, are successful examples of the application of LLMs in
bioinformatics \cite{Rose2024, Madani2020}. DrugGPT, an LLM based on the generative
pre-trained transformer (GPT) architecture \cite{Brown2020} is another example
that has shown potential in generating novel drug-like molecules having
interactions with biological targets \cite{Li2023}.

DrugGPT leverages the transformer architecture to comprehend structural
properties and structure-activity relationships. Receiving the amino
acid sequence of a given target protein, this model generates simplified
molecular input line entry system (SMILES) \cite{Weininger1988} strings of
interacting small molecules. By learning from large datasets of known
drugs and their targets, DrugGPT can propose new compounds with desired
properties by employing autoregressive algorithms for a stable and
effective training process \cite{Michailidis2013}, thus accelerating the lead
discovery phase in drug development. However, the effectiveness of
generative models in drug discovery relies heavily on the quality and
relevance of the training data \cite{Zeng2022}. Models trained on comprehensive
and accurately curated datasets are more likely to produce viable drug
candidates \cite{Kim2020}. Additionally, fine-tuning these models can enhance
their performance for predictive applications \cite{Stokes2020}.

In this study, we developed ``DrugGen'', an LLM based on the DrugGPT
architecture, finetuned using a curated dataset of approved drug-target
pairs; which is further enhanced using a policy optimization method. By
utilizing this approach, DrugGen is optimized to generate drug
candidates with optimized properties. Furthermore, we evaluated the
model's performance using custom metrics---validity, diversity, and
novelty---to comprehensively assess the quality and properties of the
generated compounds. Our results indicated that DrugGen generates
chemically sound and valid molecules in comparison with DrugGPT while
maintaining diversity and validity of generated structures. Notably,
DrugGen excels in generating molecules with higher predicted binding
affinities, increasing the likelihood of strong interactions with
biological targets. Docking simulations further demonstrated the model's
capability to accurately target binding sites and suggest new
pharmacophores. These findings highlight DrugGen's promising potential
to advance pharmaceutical research. Moreover, we proposed evaluation
metrics that can serve as objective and practical measures for comparing
future models.

\section{Results}\label{results}
\begin{figure}[h]
    \centering
    \includegraphics[width=\textwidth]{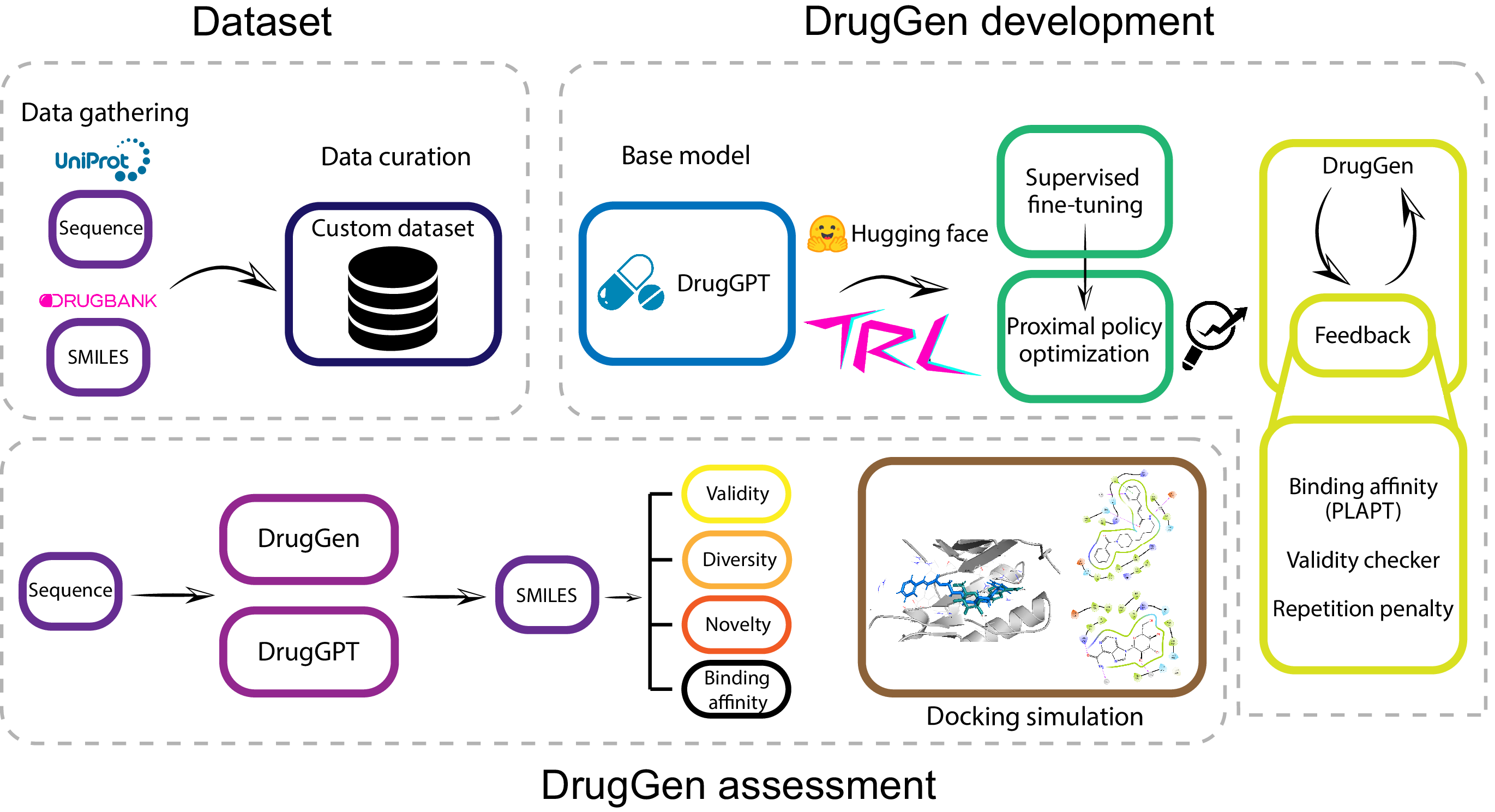}
    \caption{Schematic representation of model development and evaluation. The top section illustrates the dataset creation and the training of DrugGen through supervised fine-tuning (SFT) and proximal policy optimization (PPO) using a customized reward function. The bottom section outlines the assessment process, based on validity, diversity, novelty, and binding affinity for both DrugGen and DrugGPT, along with docking simulations for DrugGen.}
    \label{fig:study_design}
\end{figure}

In order to develop an algorithm to generate drug-like structures, we
gathered a curated dataset of approved drug-target pairs. We began by
selecting a pre-trained model and then enhanced its performance through
a two-step process. First, we employed supervised fine-tuning (SFT) on a
dataset of approved sequence-SMILES pairs to fine-tune the model. Next,
we utilized a reinforcement learning algorithm---proximal policy
optimization (PPO)---along with a customized reward system to further
optimize its performance. The final model was named DrugGen. The
schematic design of the study is illustrated in Fig. \ref{fig:study_design}.

% Figure 1

\subsection{DrugGen is effectively fine-tuned on a dataset of
approved
drug-target}\label{druggen-is-effectively-fine-tuned-on-a-dataset-of-approved-drug-target}
\begin{figure}[h]
    \centering
    \includegraphics[width=\textwidth]{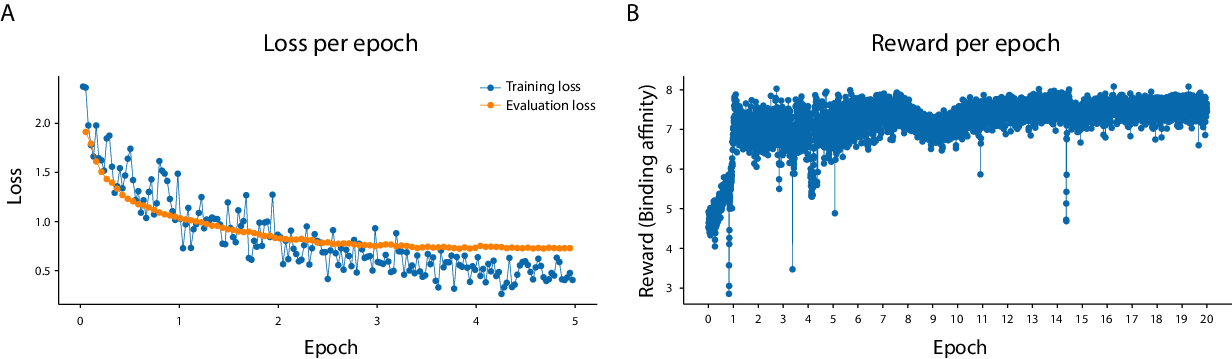}
    \caption{Training process of DrugGen. (A) Learning curve of the model during supervised fine-tuning (SFT) and (B) reward trend during proximal policy optimization (PPO).}
    \label{fig:Training}
\end{figure}

Supervised fine-tuning using the SFT trainer exhibited a steady decrease
in training and validation loss over the epochs, indicating effective
learning (Fig. \ref{fig:Training}A and \href{run:Supplementary_file_1.json}{Supplementary file 1}). After three epochs of
training, the loss of both the training and validation datasets reached
a plateau. Therefore, checkpoint number three was selected for the
second phase. In the second phase, the model was further optimized using
PPO based on the customized reward system. Over 20 epochs of
optimization, the model generated 30 unique small molecules for each
target in each epoch, ultimately reaching a plateau in the reward
diagram (Fig. \ref{fig:Training}B and \href{run:Supplementary_file_2.txt}{Supplementary file 2}).

% Figure 2

\subsection{DrugGen generates valid, diverse, and novel small
molecules}\label{druggen-generates-valid-diverse-and-novel-small-molecules}
Eight proteins were selected for models assessments which include two
targets with a high probability of association with diabetic kidney
disease (DKD) from the DisGeNet database, angiotensin-converting enzyme
(ACE) and peroxisome proliferator-activated receptor gamma (PPARG) and
six proteins without known approved drugs, i.e., galactose mutarotase
(GALM), putative fatty acid-binding protein 5-like protein 3 (FB5L3),
short-wave-sensitive opsin 1 (OPSB), nicotinamide
phosphoribosyltransferase (NAMPT), phosphoglycerate kinase 2 (PGK2), and
fatty acid-binding protein 5 (FABP5), that identified as having a high
probability of being targeted by approved small molecules through our
newly developed druggability scoring algorithm, DrugTar \cite{Borhani2024}. For
each target, 500 molecules were generated. The validity of generated
molecules was 95.45\% and 99.90\% for DrugGPT and DrugGen, respectively
(Chi-Squared, \emph{P} \textless{} 10\textsuperscript{-38},
\href{run:Supplementary_file_3.txt}{Supplementary file 3}). These molecules had an average diversity of
84.54\% {[}74.24-90.48{]} for DrugGPT and 60.32\% {[}38.89-92.80{]} for
DrugGen (\emph{U} = 358245213849, \emph{P} = 0, Fig. \ref{fig:Assessments}A and
\href{run:Supplementary_file_4.zip}{Supplementary file 4}), indicating the generation of more similar
molecules in DrugGen. These results suggest that DrugGen still generates
a wide range of structurally diverse drug candidates rather than
producing similar or redundant molecules.

To assess the novelty of generated molecules, 100 unique small molecules
were generated for each target. The validity scores for DrugGPT and
DrugGen were in agreement with previous results (95.5\% and 100\%,
respectively, Chi-Squared, \emph{P} \textless{} 10\textsuperscript{-8},
\href{run:Supplementary_file_5.txt}{Supplementary file 5}). After removing invalid structures, the novelty
scores for DrugGPT and DrugGen were 66.84\% {[}55.28-73.57{]} and
41.88\% {[}24-59.66{]}, respectively ({[}Mann--Whitney, \emph{U} = 475980, \emph{P}
\textless{} 10\textsuperscript{-80}{]}, Fig. \ref{fig:Assessments}B and \href{run:Supplementary_file_5.txt}{Supplementary file 5}), indicating that fewer novel molecules were generated in DrugGen.
These values indicate a good balance between diversity and novelty for
DrugGen.

\subsection{DrugGen generates small molecules with high affinity
for their
targets}\label{druggen-generates-small-molecules-with-high-affinity-for-their-targets}
We used two different measures to assess the binding affinity of the
generated molecules to their respective targets: PLAPT, an LLM for
predicting binding affinity, and molecular docking.

\begin{figure}[p]
    \centering
    \includegraphics[width=0.85\textwidth]{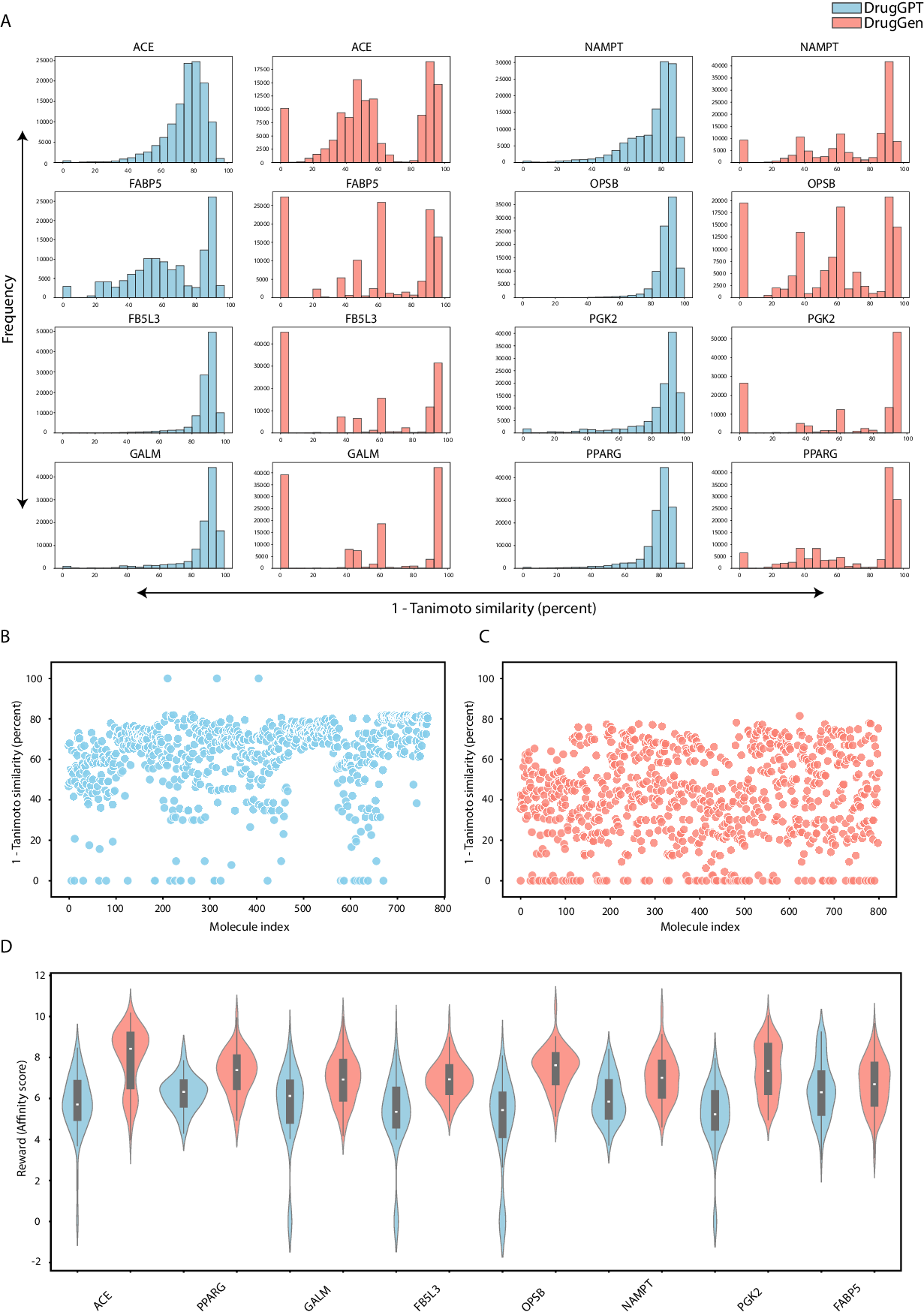}
    \caption{Comparison of molecular diversity, novelty, and binding affinity across different targets. (A) Comparison of molecular diversity distribution is shown as the frequency of the “1 – Tanimoto similarity” in percent for each target. (B) Scatter plots comparing novelty, with the “1 – Tanimoto similarity” in percent plotted against the molecular index. (C) Violin plots depicting the distribution of binding affinity for each target.}
    \label{fig:Assessments}
\end{figure}

\emph{PLAPT:} The same set of small molecules generated in novelty
assessment (100 unique small molecules for each target) were used for
assessing the quality of generated structures. Except for FABP5, DrugGen
consistently produced small molecules with significantly higher binding
affinities compared to DrugGPT ({[}7.22 {[}6.30-8.07{]} vs. 5.81
{[}4.97-6.63{]}, \emph{U} = 137934, \emph{P} \textless{}
10\textsuperscript{-85}{]}, Fig. \ref{fig:Assessments}C, Table \ref{tab:binding_affinities}, and \href{run:Supplementary_file_5.txt}{Supplementary file 5}). This finding underscores DrugGen\textquotesingle s superior
capability to generate high-quality structures.

% Figure 3

% Table 1
\begin{table}[h]
    \centering
    \caption{Statistical analysis of binding affinities of DrugGPT vs. DrugGen.}
    \label{tab:binding_affinities}
    \small
    \setlength{\abovetopsep}{-15pt}
    \begin{longtable}{@{}llccc@{}}
        \toprule
        \textbf{Targets} & \textbf{DrugGPT} & \textbf{DrugGen} & \textbf{\textit{U} statistics} & \textbf{\textit{P}} \\ 
        \midrule
        \endfirsthead
        \toprule
        \textbf{Targets} & \textbf{DrugGPT} & \textbf{DrugGen} & \textbf{\textit{U} statistics} & \textit{\textit{P}} \\ 
        \midrule
        \endhead
        \bottomrule
        \endfoot
        ACE & 5.71 {[}5.10-6.71{]} & 8.43 {[}6.65-9.06{]} & 1475 & $< 10^{-16}$* \\ 
        PPARG & 6.32 {[}5.75-6.74{]} & 7.39 {[}6.61-7.95{]} & 2208 & $< 10^{-10}$* \\ 
        GALM & 6.12 {[}4.96-6.73{]} & 6.92 {[}6.04-7.73{]} & 2767 & $< 10^{-6}$* \\ 
        FB5L3 & 5.35 {[}4.73-6.38{]} & 6.94 {[}6.36-7.48{]} & 1723 & $< 10^{-14}$* \\ 
        OPSB & 5.43 {[}4.26-6.14{]} & 7.62 {[}6.84-8.07{]} & 842 & $< 10^{-22}$* \\ 
        NAMPT & 5.84 {[}5.16-6.75{]} & 7.00 {[}6.19-7.70{]} & 2616 & $< 10^{-7}$* \\ 
        PGK2 & 5.23 {[}4.62-6.22{]} & 7.34 {[}6.36-8.52{]} & 1212 & $< 10^{-18}$* \\ 
        FABP5 & 6.30 {[}5.35-7.18{]} & 6.69 {[}5.80-7.60{]} & 4382 & 1 \\ 
    \end{longtable}
        \vspace{-3mm} All data are presented as median {[}Q1-Q3{]}. Data are compared using Mann–Whitney U test with corrections for multiple comparisons using the Bonferroni method. * \( P < 0.05 \)
\end{table}

\begin{figure}[h]
    \centering
    \includegraphics[width=\textwidth]{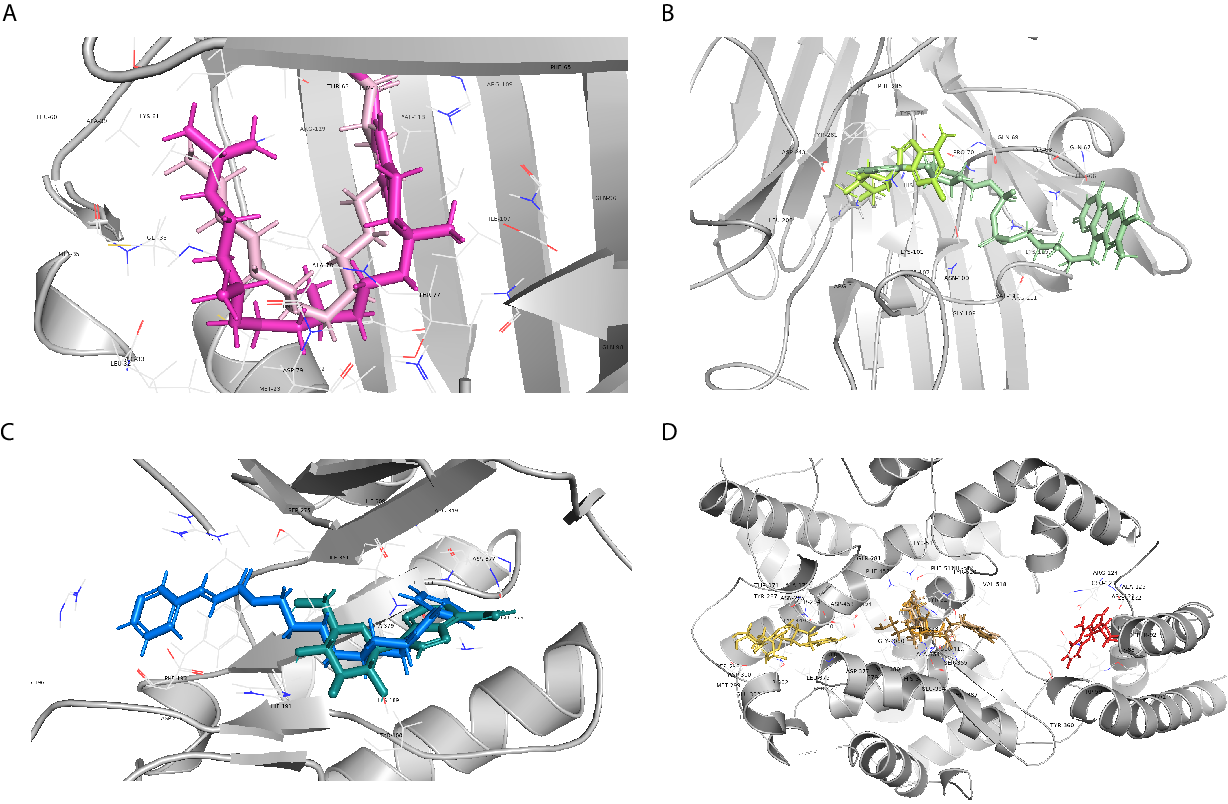}
    \caption{Visualization of ligand binding in active sites across selected targets. (A) FABP5/11 (dark pink) and Palmitic acid (light pink, reference) in the FABP5 active site. (B) GALM13 (light green) and GALM2 (lime) associated with GALM. (C) NAMPT40 (teal) and Daporinad (blue, reference) in the NAMPT active site. (D) ACE29 (red), ACE28 (yellow), ACE17 (peach), ACE14 (orange), and Lisinopril (copper, reference) associated with ACE.}
    \label{fig:Docking}
\end{figure}

\newpage
\emph{Molecular docking:} Docking simulations were performed on the
targets that had reliable protein data bank files and could be
successfully re-docked, i.e., FABP5, NAMPT, and ACE. GALM protein was
included to emphasize the model's capability to create molecules for
unexplored targets with no reference molecules. The results showed that
the generated molecules included agents with high binding affinities for
the binding site of their respective targets (Table \ref{tab:docking_scores} and \href{run:Supplementary_file_6.xlsx}{Supplementary file 6}). Except for ACE which has multiple proven binding sites with
docked molecules binding to different locations than the reference
molecule, all other docked molecules were positioned in the same binding
site as the reference in their best-docked poses (Fig. \ref{fig:Docking}).

% Table 2
\begin{table}[h]
    \setcounter{table}{1}
    \centering
    \caption{Extra precision (XP) docking scores of generated ligands.}
    \label{tab:docking_scores}
    \small
    \setlength{\abovetopsep}{-15pt}
    \begin{longtable}{@{}p{0.5\linewidth}p{0.4\linewidth}@{}}
        \toprule
        \textbf{Small molecules} & \textbf{XP GScore} \\
        \midrule
        \endfirsthead
        \toprule
        \textbf{Small Molecules} & \textbf{XP GScore} \\
        \midrule
        \endhead
        \bottomrule
        \endfoot
        NAMPT40 & -8.381 \\
        Daporinad & -8.300 \\
        NAMPT23 & -8.187 \\
        Lisinopril & -19.489 \\
        ACE17 (Enalapril) & -15.538 \\
        ACE14 (Captopril) & -9.677 \\
        ACE28 & -8.964 \\
        ACE29 & -6.405 \\
        GALM13 & -8.905 \\
        GALM2 & -7.061 \\
        GALM7 & -6.913 \\
        FABP5/11 & -9.537 \\
        FABP5/5 & -8.399 \\
        Palmitic acid & -6.177 \\
    \end{longtable}
\end{table}

% Figure 4

\begin{figure}[h]
    \centering
    \includegraphics[width=\textwidth]{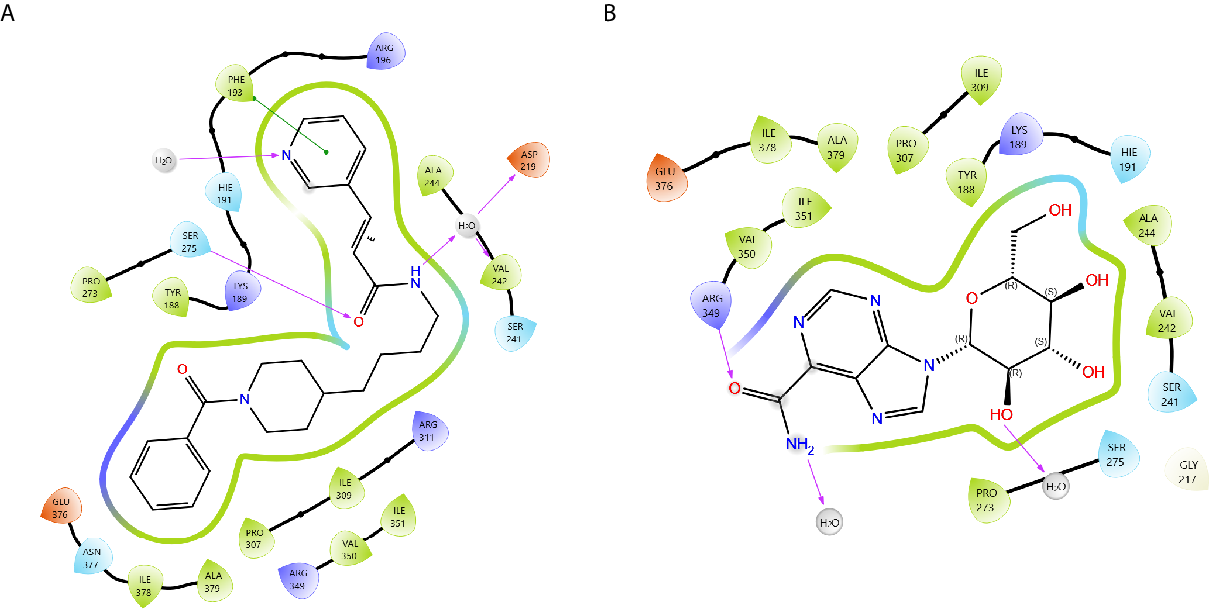}
    \caption{Comparison of pharmacophores of NAMPT inhibitors in the same active site. Daporinad (left) and NAMPT40 (right) with fundamentally different pharmacophores, both placed in the active site of NAMPT.}
    \label{fig:NAMPT}
\end{figure}

Furthermore, the model has generated molecules with better docking
scores than the reference for FABP5 (-9.537 and -8.399 vs. -6.177) and
NAMPT (-8.381 vs. -8.300). Notably, for NAMPT, the model suggested a
novel pharmacophore that occupies the same active site as the reference
molecule (Fig. \ref{fig:NAMPT}). ID cards of generated small molecules with their
related SMILES are presented in \href{run:Supplementary_file_7.txt}{Supplementary file 7}.

% Figure 5

\section{Discussion}\label{discussion}
In this study, we developed DrugGen, a large language model designed to
generate small molecules based on the desired targets as input. DrugGen
is based on a previously developed model known as DrugGPT, achieving
improvements by supervised fine-tuning on approved drugs and
reinforcement learning. These improvements aim to facilitate the
generation of novel small molecules with stronger binding affinities and
a higher probability of approval in future clinical trials. The results
indicate that DrugGen can produce high-affinity molecules with robust
docking scores, highlighting its potential to accelerate the drug
discovery process.

DrugGen is primarily based on the DrugGPT, which utilizes a GPT-2
architecture trained on datasets comprising SMILES and SMILES-protein
sequence pairs for generation of small molecules. Although DrugGPT shows
promise, it became evident that the creation of high-quality small
molecules requires more than merely ensuring ligand-target interactions.
These molecules may also exhibit essential properties, including
favorable chemical characteristics (such as stability and the absence of
cytotoxic substructures), pharmacokinetic profiles (acceptable ADME
properties---absorption, distribution, metabolism, and excretion),
pharmacodynamic attributes (efficacy and potency) \cite{Roskoski2024, Loftsson2015, DI2016}. Hence,
due to the hypothesis that approved drugs have intrinsic properties that
make them become approved \cite{Li2024}, DrugGen was fine-tuned on approved
sets of small molecules. This fine-tuning was enhanced through binding
affinity feedback from another LLM, PLAPT, resulting in improved quality
of generated molecules. Our findings demonstrate that DrugGen produces
small molecules with significantly better chemical validity and binding
affinity compared to DrugGPT while maintaining chemical diversity.

To assess the capability of DrugGen in generating high-quality
molecules, we selected eight targets. The inclusion of six targets
without known approved small molecules demonstrates
DrugGen\textquotesingle s potential to introduce novel candidates for
previously untargeted or unexplored therapeutic areas. Among the
assessed targets, generated molecules showed enhanced validity and
stronger binding affinities compared to those produced by DrugGPT. This
consistency suggests that DrugGen\textquotesingle s reinforcement
learning process effectively enhances its ability to generate potent
drug candidates. Moreover, docking simulations further confirmed the
quality of DrugGen in generating high-quality small molecules. The
comparison of docking scores between generated and reference molecules,
NAMPT40 vs. Daporinad and FABP5/11 and FABP5/5 vs. Palmitic acid, shows
that DrugGen can design molecules with predicted interactions stronger
than the known drugs. This observation highlights DrugGen's capability
to innovate beyond the existing drug design approaches. Furthermore, the
diversity of generated molecules, reflected in the wide range of docking
scores, emphasizes the model's flexibility in producing varied chemical
structures. Additionally, in the case of NAMPT, the model generated one
structure with a strong docking score possessing a pharmacophore very
different from that of the reference molecules, meaning that core drug
structure was dissimilar to the reference molecule. This structure
occupied the same binding site as the reference molecule, which is a
potentially new pharmacophore for this target. In addition to these
improvements, in the process of reinforcement learning, penalties were
applied for generating repetitive structures, resulting in a diverse and
valid set of molecules whilst retaining the possibility of regenerating
approved drugs in the case of drug repurposing \cite{Kulkarni2023}. Thus, DrugGen
demonstrates applicability in both \emph{de novo} drug design and
repurposing efforts.

Despite these achievements, our study has some limitations that should
be considered in future research. Variability in binding affinity
results across assessed targets was observed. For instance, FABP5's
performance improvement was less pronounced compared with others. This
might suggest that with certain target classes or protein sequences,
unique challenges emerge for our model, requiring additional fine-tuning
or alternative strategies for further optimization. In addition, DrugGen
cannot target a specific binding site, as can be seen in the case of
ACE, which has multiple binding sites \cite{Cozier2021}. Ligand prediction using
the DrugGen model led to molecules with fairly strong ligand binding to
different binding sites; however, this may not be desirable in some
cases. The existing reward function relies on an affinity-predictor deep
learning model that has inherent accuracy and specificity limitations
due to the limitations of the databases and input representation, which
could be addressed in future works. Our model is primarily focused on
predicting novel cores and structures for targets with limited bioactive
molecules, thus it does not generate fully optimized structures. These
predicted structures should undergo structural manipulation for
structural optimization to better fit the active site of the target.
Future improvements will involve incorporating active site interactions
into the reward system to enhance structural accuracy. Finally, the
reliance on \emph{in silico} validation, while useful, needs to be
complemented with experimental validation to confirm the practical
efficacy and safety of the generated molecules.

In conclusion, DrugGen represents a powerful tool for early-stage drug
discovery, with the potential to significantly accelerate the process of
identifying novel lead compounds. With further refinement and
integration with experimental validation, DrugGen could become an
integral part of future drug discovery pipelines, contributing to the
development of new therapeutics across a wide range of diseases.

\section{Materials and Methods}\label{materials-and-methods}
\subsection{Dataset Preparation}\label{dataset-preparation}
A dataset of small molecules, each approved by at least one regulatory
body, was collected to enhance the safety and relevance of the generated
molecules. First, 1710 small molecules from the DrugBank database
(version: 2023-01-04) were retrieved \cite{Wishart2006}, 117 of which were labeled
as withdrawn. After initial assessments of withdrawn drugs by a
physician (Ali Motahharynia) and a pharmacist (Mahsa Sheikholeslami),
consensus was reached to omit 50 entries due to safety concerns.
Consequently, 1660 approved small molecules and their respective targets
were selected. From the total of 2116 approved drug targets, retrieved
from DrugBank database, 27 were not present in the UniProt database \cite{UniProt}. After further assessment, these 27 proteins were replaced
manually with equivalently reviewed UniProt IDs, identical protein
names, or by basic local alignment search tools (BLAST) \cite{Altschul1990}. The
protein with UniProt ID ``Q5JXX5'' was deleted from the UniProt database
and therefore, omitted from the collected dataset as well. Finally, 1660
small molecules and 2093 related protein targets were selected.
Available SMILEs (1634) were retrieved from DrugBank, ChEMBL \cite{Zdrazil2023},
and ZINC20 databases \cite{Irwin2020}. Protein sequences were retrieved from the
UniProt database.

\subsection{Data Preprocessing}\label{data-preprocessing}
Similar to the structure used by DrugGPT, the small molecules and target
sequences were merged into the pair of a string consisting of protein
sequence and SMILES in the following format:
``\textless\textbar startoftext\textbar\textgreater{} +
\textless P\textgreater{} + target protein sequence +
\textless L\textgreater{} + SMILES +
\textless\textbar endoftext\textbar\textgreater." To ensure the
compatibility of this input format with the original model, the
resulting strings were tokenized using the trained DrugGPT's byte-pair
encoding (BPE) tokenizer (53083 tokens). The strings were padded to the
maximum length of 768, and longer strings were truncated. The
``\textless\textbar startoftext\textbar\textgreater'',
``\textless\textbar endoftext\textbar\textgreater'', and
``\textless PAD\textgreater'' were defined as special tokens.

\subsection{DrugGen Development Overview}\label{drugGen-development-overview}
Using the supervised fine-tuning (SFT) trainer module from the
transformer reinforcement learning (TRL) library (version: 0.9.4)
\cite{trl2024}, the original DrugGPT model was finetuned on our dataset.
Afterward, reinforcement learning was applied to further improve the
model. For this purpose, a Tesla V100 GPU with 32 GB of VRAM, 64 GB of
RAM, and a 4-core CPU were utilized for both phases, i.e., SFT and
reinforcement learning using a PPO trainer.

\subsubsection{Supervised
Fine-tuning}\label{supervised-fine-tuning}
The training dataset consisted of 9398 strings. The base model was
trained using the SFT trainer class for five epochs with the following
configuration: Learning rate: 5e-4, batch size: 8, warmup steps (linear
warmup strategy): 100, and eval steps: 50. AdamW optimizer with a
learning rate of 5e-4 and epsilon value of 1e-8 was used for optimizing
the model parameters. The model\textquotesingle s performance on the
training and validation sets (ratio of 8:2) was evaluated using the
cross-entropy loss function during the training phase.

\subsubsection{Proximal Policy
Optimization}\label{proximal-policy-optimization}
Hugging Face's PPO Trainer, which is based on OpenAI's original method
for ``Summarize from Feedback'' \cite{summarize2024} was used in this study. PPO is
a reinforcement learning algorithm that improves the policy by taking
small steps during optimization, avoiding overly large updates that
could lead to instability. The key formula used in PPO is:

\begin{equation}
L^{CLIP}(\theta) = E_{t}\left[ \min(r_{t}(\theta) A_{t}, \text{clip}(r_{t}(\theta), 1 - \epsilon, 1 + \epsilon) A_{t}) \right]
\end{equation}

In this equation, \( L^{CLIP}(\theta) \) represents the clipped objective function that PPO aims to optimize during training. The expectation \( E_{t} \) denotes the average over time steps \( t \), capturing the overall performance of the policy. The term \( r_{t}(\theta) \) is the probability ratio of taking action \( a_{t} \) under the new policy compared to the old policy, defined as
\(
r_{t}(\theta) = \frac{\pi_{\theta}(a_{t} \mid s_{t})}{\pi_{\theta_{\text{old}}}(a_{t} \mid s_{t})}.
\)
The advantage estimate \( A_{t} \) quantifies the relative value of the action taken in relation to the expected value of the policy.
The clipping function, \(clip(r_{t}(\theta),1 - \epsilon,1 + \epsilon)\),
restricts the ratio to a defined range, preventing large updates to the
policy that could destabilize training. This formulation allows PPO to
balance exploration and stability, enabling effective policy updates
while minimizing the risk of performance degradation. There are three
main phases in training a model with PPO. First, the language model
generates a response based on an input query in a phase called the
rollout phase. In our study, the queries were protein sequences, and the
generated responses were SMILES strings. Then in the evaluation phase,
the generated molecules were assessed with a custom model that predicts
binding affinity. Finally, the log probabilities of the tokens in the
generated SMILES sequences were calculated based on the query/response
pairs. This step is also known as the optimization phase. Additionally,
to maintain the generated responses within a reasonable range from the
reference language model, a reward signal was introduced in the form of
the Kullback-Leibler (KL) divergence between the two outputs. This
additional signal ensures that the new responses do not deviate too far
from the original model\textquotesingle s outputs. Thus, PPO was applied
to train the active language model.

In our study, the rollout section had the following generation
parameters: ``do\_sample'': True, ``top\_k'': 9, ``top\_p'': 0.9,
``max\_length'': 1024, and ``num\_return\_sequences'': 10. In each
epoch, generation was continued until 30 unique small molecules were
generated for each target. Keeping initial model's structure in mind,
the dataset was filtered based on the length of each protein sequence.
After creating the prompts according to the specified format, i.e.,
``\textless\textbar startoftext\textbar\textgreater{} +
\textless P\textgreater{} + target protein sequence +
\textless L\textgreater'', prompts with a tensor size greater than 768
were omitted, resulting in 2053 proteins (98.09\% of the initial
dataset).

The PPO trainer configuration included: ``mini\_batch\_size'': 8,
``batch\_size'': 240, and ``learning\_rate'': 1.41e-5. Score scaling and
normalization were handled with the PPO trainer's built-in functions.

\subsubsection{Reward Function}\label{reward-function}
\emph{PLAPT:} PLAPT, a cutting-edge model designed to predict binding
affinities with remarkable accuracy was used as a reward function. PLAPT
leverages transfer learning from pre-trained transformers, ProtBERT and
ChemBERTa, to process one-dimensional protein and ligand sequences,
utilizing a branching neural network architecture for the integration of
features and estimation of binding affinities. The superior performance
of PLAPT has been validated across multiple datasets, where it achieved
state-of-the-art results \cite{Rose2024}. The affinities of the generated
structures with their respective targets were evaluated using PLAPT's
neg\_log10\_affinity\_M output.

\emph{Customized invalid structure assessor:} We developed a customized
algorithm using RDKit library (version: 2023.9.5) \cite{rdkit2024} to assess
invalid structure, where specific checks were performed to identify
potential issues such as atom count, valence errors, and parsing errors.
Invalid structures, including those with fewer than two atoms, incorrect
valence states, or parsing failures were flagged and penalized
accordingly. To promote the generation of valid molecules, a reward
value of 0 was assigned to any invalid SMILES structures. These reward
systems provide a rigorous scoring system for model development.

To further shift the model toward generating novel molecules, a
multiplicative penalty was applied to the reward score when a generated
SMILES string matched a molecule already present in the approved SMILES
dataset. Specifically, the reward was multiplied by 0.7 for such
occurrences, to retain a balance between generating new structures as
well as repurposing approved drugs.

\subsubsection{DrugGen Assessment}\label{druggen-assessment}
To evaluate the performance of DrugGen, several metrics were employed to
measure its efficacy in generating viable and high-affinity drug
candidates. For this purpose, eight targets consisting of two DKD
targets with the highest score in DisGeNet database (version 3.12.1)
\cite{DisGeNet2017}, i.e., ``ACE'' and ``PPARG'' and six targets without any known
approved small molecules for them were selected. The selection of these
six targets was according to our recent study ``DrugTar Improves
Druggability Prediction by Integrating Large Language Models and Gene
Ontologies'' \cite{Borhani2024}. According to this study, 6 out of the 10 most
probable proteins for future targets were selected. The selected targets
are ``GALM'', ``FB5L3'', ``OPSB'', ``NAMPT'', ``PGK2'', and ``FABP5''.
The generative quality of DrugGPT and DrugGen in terms of validity,
diversity, novelty, and binding affinity was assessed. Additionally, we
performed in silico validation of the molecules generated by DrugGen
using a rigorous docking method.

\subsubsubsection{Validity Assessment}\label{validity-assessment}
The validity of the generated molecules was evaluated using the
previously mentioned customized invalid structure assessor. The
percentage of valid to total generation was reported as models'
capability to construct valid structures.

\subsubsubsection{Diversity Assessment}\label{diversity-assessment}
To assess the diversity of the generated molecules, 500 ligands were
generated for each target by DrugGPT and DrugGen. The diversity of the
generated molecules was quantitatively assessed using the Tanimoto
similarity index \cite{Bajusz2015}. The diversity evaluation process involved the
following steps: First, each generated molecule was converted to its
corresponding molecular fingerprint using Morgan fingerprints (size =
2048 bits, radius = 2) \cite{Morgan1965}. For each molecule, pairwise Tanimoto
similarities were calculated between all possible pairs of fingerprints,
and the average value was calculated. Thus, the diversity of the
generated set was determined as the ``1 - average of Tanimoto
similarity'' within a generated batch. The distribution of diversity for
each target was plotted. The invalid structures were not involved in
diversity assessments. Statistical analyses were performed using
Mann--Whitney \emph{U} test.

\subsubsubsection{Novelty Assessment}\label{novelty-assessment}
For each target, a set of 100 unique molecules was generated by DrugGPT
and DrugGen. The novelty of the generated molecules was evaluated by
comparing them to a dataset of approved drugs. After converting the
molecules into Morgan fingerprints, the similarity of each generated
molecule to the approved drugs was calculated using Tanimoto similarity
index, retaining only the maximum similarity value. The novelty was
reported as the ``1 - max\_Tanimoto similarity''. The invalid structures
were not included in the novelty assessments. Statistical analyses were
performed using Mann--Whitney \emph{U} test.

\subsubsubsection{PLAPT Binding Affinity Assessment}\label{plapt-binding-affinity-assessment}
The same set of molecules generated during the novelty assessment was
used to evaluate the binding affinities of the compounds produced by
DrugGPT and DrugGen. The invalid structures were involved in the binding
affinity assessments. Statistical analysis was conducted using the
Mann--Whitney \emph{U} test, and corrections for multiple comparisons
were applied using the Bonferroni method.

\subsubsubsection{Molecular Docking}\label{molecular-docking}
Molecular docking was conducted for selected targets with available
protein data bank (PDB) structures, specifically ACE, NAMPT, GALM, and
FABP5. A set of 100 newly generated molecules, following duplicate
removal, were docked into the crystal structures of ACE (PDB ID: 1o86),
NAMPT (PDB ID: 2gvj), GALM (PDB ID: 1snz), and FABP5 (PDB ID: 1b56).
Overall, blind docking \cite{Hassan2017} was employed for all 122 generated
molecules and their references to thoroughly search the entire protein
surface for the most favorable active site (\href{run:Supplementary_file_6.xlsx}{Supplementary file 6} and \href{run:Supplementary_file_7.txt}{Supplementary file 7}).
The reference ligands used were Lisinopril for ACE and Palmitic acid for
FABP5, both of which were bound in the active site. For NAMPT,
Daporinad, a molecule currently in phase 2 clinical trials, served as
the highest available reference. In the case of GALM, no reference
ligand was found. The retrieved PDB files were prepared using the
protein preparation wizard \cite{Madhavi2013} available in the Schrödinger suite,
ensuring the addition of missing hydrogens, assignment of appropriate
charge states at physiological pH, and reconstruction of incomplete side
chains and rings. LigPrep \cite{schrodinger2024} with the OPLS4 force field \cite{Lu2021}
was employed to generate all possible stereoisomers and ionization
states at pH 7.4±0.5. The prepared structures were used for docking.

Docking simulations were performed using the GLIDE program cite{Friesner2004}.
Ligands were docked using the extra precision (XP) protocol. Ligands
were allowed full flexibility during the docking process, while the
protein was held rigid. The information of the grid boxes is summarized
in Table \ref{tab:gridbox}.

% Table 3: Gridbox Generation Properties
\begin{table}[h]
    \setcounter{table}{2} 
    \centering
    \caption{Gridbox generation properties for performing blind docking.}
    \label{tab:gridbox}
    \small % Reduces font size
    \setlength{\abovetopsep}{-15pt} 
    \begin{longtable}{@{}p{2.8cm} p{1.5cm} p{1.5cm} p{1.5cm} p{1.2cm} p{1.1cm} p{1.1cm} p{1.1cm} p{1cm} p{1.2cm}@{}}
        \toprule
        \textbf{Target} & \textbf{ligxrange} & \textbf{ligyrange} & \textbf{ligzrange} & \textbf{xcent} & \textbf{xrange} & \textbf{ycent} & \textbf{yrange} & \textbf{zcent} & \textbf{zrange} \\
        \midrule
        \endfirsthead
        \toprule
        \textbf{Target} & \textbf{ligxrange} & \textbf{ligyrange} & \textbf{ligzrange} & \textbf{xcent} & \textbf{xrange} & \textbf{ycent} & \textbf{yrange} & \textbf{zcent} & \textbf{zrange} \\
        \midrule
        \endhead
        \bottomrule
        \endfoot
        2GVJ - NAMPT & 40 & 40 & 40 & 14.616 & 76 & -7.569 & 76 & 14.046 & 76 \\
        1O86 - ACE & 40 & 40 & 40 & 40.657 & 76 & 37.169 & 76 & 43.527 & 76 \\
        1SNZ - GALM & 40 & 40 & 40 & -10.433 & 58 & 5.656 & 58 & 50.197 & 58 \\
        1B56 - FABP5 & 30 & 30 & 30 & 49.969 & 52 & 22.227 & 52 & 32.492 & 52 \\
    \end{longtable}
\end{table}
The GLIDE XP scoring function was used to evaluate docking poses.
Negative values of the GLIDE score (XP GScore) were reported for
readability. The robustness of the docking procedures was validated by
redocking the reference ligands into their respective binding sites. The
computed root-mean-squared deviation (RMSD) values were 0.7233Å,
0.2961Å, and 2.0119Å for ACE, NAMPT, and FABP5, respectively, confirming
the reliability of the docking protocol.
\newpage

\textbf{Data availability}

All data generated or analyzed during this study are included in the
manuscript and supporting files. The sequence-SMILES dataset of approved
drug-target pairs used in this study is publicly available at
``alimotahharynia/approved\_drug\_target'' from Hugging Face
(\url{https://huggingface.co/datasets/alimotahharynia/approved_drug_target}).

\textbf{Code availability}

The checkpoints, code for generating small molecules, and customized
validity small assessor are publicly available at
\url{https://huggingface.co/alimotahharynia/DrugGen} and
\url{https://github.com/mahsasheikh/DrugGen}.

\textbf{Acknowledgment}

We sincerely thank Dr. Mehdi Rahmani for his invaluable assistance with
technical and software issues related to training our model on the
cluster servers.

\textbf{Funding}

No funding was received for this study or its publication.

\textbf{Competing interest}

The authors declare no competing interests.

\textbf{Authors contribution}

Conceptualization: M.S, Y.G, M.I, A.M. Dataset preparation: M.S, A.M.
Model development: M.S, N.M, M.I, A.M. Statistical analysis: M.S, N.M,
A.M. In silico validation: M.S, A.F. Data interpretation: All authors.
Drafting original manuscript: M.S, N.M. Revising the manuscript: Y.G,
A.F, M.I, A.M. All the authors have read and approved the final version
for publication and agreed to be responsible for the integrity of the
study.

\end{document}